 \documentclass[twocolumn,prl,aps,psfig,showpacs,preprintnumbers,superscriptaddress]{revtex4}
\usepackage{graphicx}
\usepackage{epstopdf}
\usepackage{float}
\usepackage{color}
\usepackage{amsmath}

\begin{document}

\title{NMR studies of the  incommensurate helical antiferromagnet  EuCo$_2$P$_2$ : determination of the antiferromagnetic propagation vector}
\author{Nonoka Higa}
\affiliation{Ames Laboratory, U.S. DOE, and Department of Physics and Astronomy, Iowa State University, Ames, Iowa 50011, USA}
\affiliation{Department of Physics and Earth Sciences, Faculty of Science, University of the Ryukyus, Okinawa 903-0213, Japan}
\author{Qing-Ping Ding}
\affiliation{Ames Laboratory, U.S. DOE, and Department of Physics and Astronomy, Iowa State University, Ames, Iowa 50011, USA}
\author{Mamoru Yogi}
\affiliation{Department of Physics and Earth Sciences, Faculty of Science, University of the Ryukyus, Okinawa 903-0213, Japan}
\author{N.~S.~Sangeetha}
\affiliation{Ames Laboratory, U.S. DOE, and Department of Physics and Astronomy, Iowa State University, Ames, Iowa 50011, USA}
\author{Masato Hedo}
\affiliation{Department of Physics and Earth Sciences, Faculty of Science, University of the Ryukyus, Okinawa 903-0213, Japan}
\author{Takao Nakama}
\affiliation{Department of Physics and Earth Sciences, Faculty of Science, University of the Ryukyus, Okinawa 903-0213, Japan}
\author{Yoshichika \=Onuki}
\affiliation{Department of Physics and Earth Sciences, Faculty of Science, University of the Ryukyus, Okinawa 903-0213, Japan}
\author{D.~C.~Johnston}
\affiliation{Ames Laboratory, U.S. DOE, and Department of Physics and Astronomy, Iowa State University, Ames, Iowa 50011, USA}
\author{Yuji Furukawa}
\affiliation{Ames Laboratory, U.S. DOE, and Department of Physics and Astronomy, Iowa State University, Ames, Iowa 50011, USA}

\date{\today}

\begin{abstract} 

    Recently Ding {\it et~al.} [Phys. Rev. B {\bf 95}, 184404 (2017)] reported that their nuclear magnetic resonance (NMR) study on EuCo$_2$As$_2$ successfully characterized the antiferromagnetic (AFM) propagation vector of the incommensurate helix AFM state, showing that NMR is a unique tool for determination of  the spin structures in incommensurate helical AFMs. 
    Motivated by this work, we have carried out  $^{153}$Eu, $^{31}$P and $^{59}$Co NMR measurements on the helical antiferromagnet EuCo$_2$P$_2$ with an  AFM ordering temperature $T_{\rm N}$ = 66.5 K. 
  An incommensurate helical AFM structure  was clearly confirmed by $^{153}$Eu and $^{31}$P NMR spectra on single crystalline  EuCo$_2$P$_2$ in zero magnetic field at 1.6 K and its external magnetic field dependence. 
    Furthermore, based on $^{59}$Co NMR data in both the paramagnetic and the incommensurate AFM states, we have determined the model-independent value of the AFM propagation vector {\bf k} = (0, 0, 0.73 $\pm$ 0.09)2$\pi$/$c$ where $c$ is the $c$-axis lattice parameter. 
 The temperature dependence of  {\bf k}  is also discussed.

\end{abstract}

 \pacs{75.25.-j, 75.50.Ee, 76.60.-k}
\maketitle

 \section{I. Introduction} 
    A great deal of attention in magnetism has been given to systems with geometric frustration \cite{Lacroix,HTD}.
   Such competing  magnetic interactions between localized spins often result in noncollinear magnetic structures, such as incommensurate helical magnetic structures.
   Usually,  such noncollinear magnetic structures are determined by using neutron diffraction (ND) measurements.

   Very recently nuclear magnetic resonance (NMR)  has been shown to be another unique tool to determine spin structures in incommensurate helical antiferromagnets (AFM) \cite{Ding2017}.
    By performing $^{153}$Eu, $^{75}$As and $^{59}$Co NMR measurements on the incommensurate helical AFM EuCo$_2$As$_2$ with an ordering temperature $T_{\rm N}$ = 45 K,  the AFM propagation vector of the incommensurate helical AFM state was successfully determined \cite{Ding2017}. 
     Such an NMR approach can be used to characterize the magnetic structure in other possible helical magnets such as the isostructural metallic compound EuCo$_2$P$_2$ \cite{Reehuis1992} and EuCu$_{1.82}$Sb$_2$ \cite{Anand2015}. 
      


     Motivated by the above NMR work, we have carried out NMR measurements to characterize  EuCo$_2$P$_2$ with the body-centered tetragonal ThCr$_2$Si$_2$-type structure  which is reported to exhibit an incommensurate helical AFM ground state below $T_{\rm N}$ = 66.5 K \cite{Marchand1978,Morsen1988,Reehuis1992,Nakama2010,Sangeetha2016}.
     The neutron diffraction  measurements \cite{Reehuis1992} on  EuCo$_2$P$_2$ report that the Eu ordered moment  at 15 K is 6.9 $\mu_{\rm B}$/Eu, where $\mu_{\rm B}$ is the Bohr magneton, consistent with  Eu$^{2+}$ ($J = S = $ 7/2) and spectroscopic splitting factor $g$ = 2. 
The magnetic structure is the same as that in EuCo$_2$As$_2$ where the Eu ordered moments are aligned ferromagnetically in the $ab$ plane with the helix axis along the $c$ axis \cite{Reehuis1992}.
      The AFM propagation vector {\bf k} = (0, 0, 0.852)2$\pi$/$c$ at 15~K was determined by the ND measurements \cite{Reehuis1992}, where $c$ is the $c$-axis  lattice parameter.  
      The similar value of  {\bf k} = (0, 0, 0.88)2$\pi$/$c$ at $T$ = 0~K  (Ref. \cite{Sangeetha2016}) was also obtained by the analysis of $\chi$ data on a single crystal below $T_{\rm N}$ using molecular field theory which has been recently formulated to apply to  planar noncollinear Heisenberg antiferromagnets \cite{Johnston2012, Gotesch2014,Johnston2015}.
     It is important to independently determine the AFM propagation vector {\bf k} by using the NMR technique.
     
    Another interesting feature of metallic EuCo$_2$P$_2$ is a change in magnetic properties observed under high pressure \cite{Chefki1998}. 
    With the application of pressure, EuCo$_2$P$_2$ exhibits a first order tetragonal to collapsed-tetragonal transition and associated valence transition from Eu$^{2+}$ 
to nonmagnetic Eu$^{3+}$  ($J$ = 0) at 3.1 GPa.  
    Below 3.1 GPa, the AFM ordering  originates from Eu  4$f$ local moments where Co moments are considered to be not involved in the magnetic ordering.
     On the  other hand, above 3.1 GPa,  the change in the Eu valence from 2+ to 3+ leads to the appearance of itinerant 3$d$ magnetic ordering below 260 K. 
     Thus it is interesting and important to characterize the magnetic and electronic states of each ion in EuCo$_2$P$_2$ from a microscopic point of view. 

\begin{figure}[t]
\includegraphics[width=7.0cm]{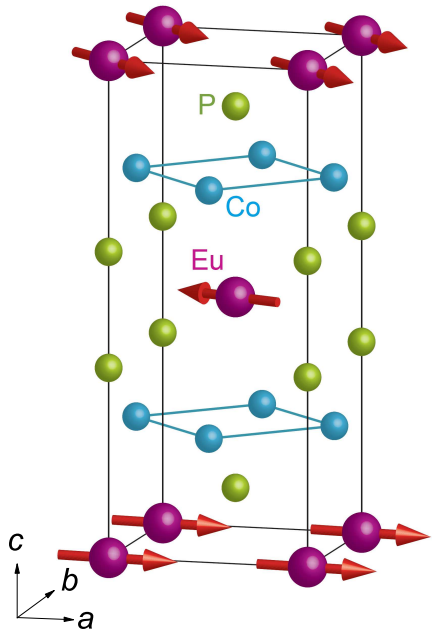} 
\caption{(Color online)  Crystal and magnetic structures of EuCo$_2$P$_2$.  
The arrows on the Eu atoms indicate the directions of the Eu ordered  moments in the incommensurate helical antiferromagnetic state.
 }
\label{fig:Structure}
\end{figure}

    In this paper, we report NMR results on  EuCo$_2$P$_2$, where we succeeded in observing NMR signals from all three $^{153}$Eu, $^{59}$Co and $^{31}$P nuclei, focusing  our attention on the spin structure in the incommensurate helical AFM state with the aim of obtaining better understandings of the local magnetic and electronic properties of the three ions in the AFM and paramagnetic states at ambient pressure.  
     From the external field dependence of $^{153}$Eu and $^{31}$P NMR spectra at 1.6 K, below $T_{\rm N}$ = 66.5 K an incommensurate helical AFM state shown in  Fig.\ \ref{fig:Structure}  was clearly evidenced in  EuCo$_2$P$_2$.
    Furthermore, the AFM propagation vector characterizing the helical AFM state is determined to be {\bf k} = (0, 0, 0.73 $\pm$ 0.09)2$\pi$/$c$  from the internal magnetic field at the Co site obtained by $^{59}$Co NMR under zero magnetic field.
   The estimated value is slightly smaller than those reported from  the neutron diffraction and magnetic susceptibility measurements \cite{Reehuis1992,Sangeetha2016}.
   $^{59}$Co NMR revealed that no magnetic ordering of the Co $3d$ electron spins occurs in the helix AFM state, evidencing that the magnetism in EuCo$_2$P$_2$ originates from only the Eu spins. 
    The temperature dependence of the Eu ordered  moments determined by the internal magnetic induction at the P site $B_{\rm int}^{\rm P}$ can be well reproduced  by the Brillouin function with $J$ =$S$ = 7/2, confirming that the magnetic state of the Eu$^{2+}$ ions is well explained by the local moment picture although the system is metallic.
    Our NMR study shows again that NMR is a powerful tool for determination of  the spin structure in incommensurate helical AFMs.

\begin{figure}[tb]
\includegraphics[width=7.0cm]{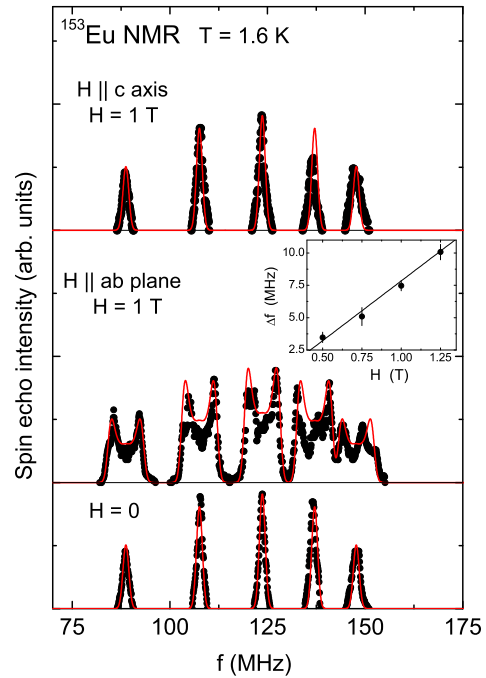} 
\caption{(Color online) $^{153}$Eu-NMR spectra at $T$ = 1.6 K in the AFM state for single-crystalline EuCo$_2$P$_2$ in $H$ = 0  (bottom panel),  $H$ = 1~T parallel to the $ab$ plane (middle panel) and parallel to the $c$ axis (top panel). The red lines constitute the calculated $^{153}$Eu NMR spectrum. 
The inset shows the external magnetic field dependence of the amount of the splitting of the central transition line ($\Delta$$f$).  The solid line is the expected $H$ dependence of $\Delta$$f$ described in the text. 
 }
\label{fig:EuNMR}
\end{figure}

 \section{II. Experimental}

   A single crystal  ($1\times1\times0.3$ mm$^3$) of EuCo$_2$P$_2$ for the NMR measurements was grown using Sn  flux \cite{Marchand1978,Sangeetha2016}.
   NMR measurements of $^{153}$Eu ($I$ = $\frac{5}{2}$, $\frac{\gamma_{\rm N}}{2\pi}$ = 4.632 MHz/T, $Q=$ 2.49 barns),   $^{59}$Co ($I$ = $\frac{7}{2}$,  $\frac{\gamma_{\rm N}}{2\pi}$ = 10.03 MHz/T, $Q=$ 0.4 barns), and $^{31}$P ($I$ = $\frac{1}{2}$, $\frac{\gamma_{\rm N}}{2\pi}$ = 17.235 MHz/T) nuclei were conducted using a homemade phase-coherent spin-echo pulse spectrometer. 
   In the AFM state, $^{153}$Eu, $^{31}$P and $^{59}$Co NMR spectra in zero and nonzero magnetic fields $H$ were measured in steps of frequency $f$ by measuring the intensity of the Hahn spin echo. 
   We have used a single crystal for  $^{153}$Eu, $^{31}$P NMR spectrum measurements at the lowest temperature of 1.6 K.  
      Above  $T$ = 1.6~K,  we performed our measurements using powdered single crystals as intensities of NMR signals with the single crystals were too weak to perform  the measurements  at higher temperatures. 
    In the paramagnetic (PM) state,  $^{59}$Co and $^{31}$P NMR spectra were obtained by sweeping the magnetic field at $f$ = 51.2 MHz.

  \section{III. Results and discussion}
  \subsection{A. $^{153}$Eu NMR spectrum }   

      The bottom panel of Fig.\ \ref{fig:EuNMR} shows the $^{153}$Eu NMR spectrum in the AFM state for EuCo$_2$P$_2$ (single crystal)  measured in zero magnetic field at a temperature $T$ = 1.6 K. 
       An almost identical $^{153}$Eu NMR spectrum was observed on a powder sample EuCo$_2$P$_2$ at 1.6 K (not shown). 
        The observed spectrum is well reproduced by the following nuclear spin Hamiltonian for the case that the Zeeman interaction is much greater than the quadrupole interaction, which produces a spectrum with a central transition line flanked by two satellite peaks on both sides for $I$ = 5/2,  
\begin{equation}
   {\cal H} = {\cal -}\gamma\hbar{\bf I}\cdot{\bf B_{\rm int}}+ \frac{h \nu_{\rm Q}}{6} [3I_{z}^{2}-I(I+1) + \frac{1}{2}\eta(I_+^2 +I_-^2)], 
\end{equation}
where $B_{\rm int}$ is the internal magnetic induction at the Eu site, $h$ is Planck's constant, and $\nu_{\rm Q}$ is the nuclear quadrupole frequency defined by $\nu_{\rm Q} = 3e^2QV_{ZZ}/2I(2I-1)h$  $(=3e^2QV_{ZZ}/20h$ for $I$ = 5/2)  where $Q$ is the electric quadrupole moment of the Eu nucleus, $V_{ZZ}$ is the electric field gradient (EFG) at the Eu site, and $\eta$ is the asymmetry parameter of the EFG \cite{Slichter_book}.
    Since the Eu site in EuCo$_2$P$_2$ has a tetragonal local symmetry (4$/mmm$), $\eta$ is zero. 
   The red line shown at the bottom panel of Fig.\ \ref{fig:EuNMR} is the calculated spectrum for $^{153}$Eu zero-field NMR (ZFNMR) using the parameters $|B_{\rm int}^{\rm Eu}|$  = 25.75(2) T (= 119.3 MHz), $\nu_{\rm Q}$ = 30.2(2)  MHz and $\theta = 90^\circ$. 
   Here $\theta$ represents the angle between $B_{\rm int}^{\rm Eu}$  and the principal axis of the EFG tensor at the Eu sites. 
 
   Since the principal axis of the EFG  at the Eu site is parallel to the $c$ axis due to the local symmetry \cite{Yogi2013,Ding2017}, $\theta = 90^\circ$ indicates that $B_{\rm int}^{\rm Eu}$ is perpendicular  to the $c$ axis.
   This is similar to the case of the Eu nuclei in EuCo$_2$As$_2$ and EuGa$_4$ with the same ThCr$_2$Si$_2$-type  crystal structure in which the similar values of $B_{\rm int}^{\rm Eu}$ = 27.5~T and 27.08~T and  $\nu_{\rm Q}$ = 30.6~MHz and 30.5~MHz for $^{153}$Eu, respectively,  have been reported \cite{Yogi2013,Ding2017}.
         $B_{\rm int}^{\rm Eu}$ is proportional to $A_{\rm hf}$$<$$\mu$$>$ where $A_{\rm hf}$ is the  hyperfine coupling constant and $<$$\mu$$>$ is the ordered Eu magnetic moment. 
      The hyperfine field at the Eu sites mainly originates from core polarization from 4$f$ electrons and is oriented in a direction opposite to that of the Eu moment \cite{Freeman1965}.
      For $|$$B_{\rm int}^{\rm Eu}$$|$ = 25.75(2)~T and the reported AFM ordered moment $<$$\mu$$>$  = 6.9(1)~$\mu_{\rm B}$/Eu from ND \cite{Reehuis1992},  $A_{\rm hf}$ is estimated to be  $-$3.73~T/$\mu_{\rm B}$ where the sign is reasonably assumed to be negative due to the core-polarization mechanism. 
       The estimated $A_{\rm hf}$ is very close to  $-$3.78~T/$\mu_{\rm B}$ for the case of  EuCo$_2$As$_2$ \cite{Ding2017} and is not far from  the core-polarization hyperfine coupling constant $-$4.5~T/$\mu_{\rm B}$ estimated for Eu$^{2+}$ ions \cite{Freeman1965}. 
      The small difference could be explained by a positive hyperfine coupling contribution due to conduction electrons which cancels part of the negative core polarization field as has been pointed out in the case of EuCo$_2$As$_2$ (Ref. \cite{Ding2017}).  

   The direction of $B_{\rm int}^{\rm Eu}$  is also directly confirmed by $^{153}$Eu NMR spectrum measurements on the single crystal in nonzero $H$. 
   When $H$ is applied along the $c$ axis, almost no change of the $^{153}$Eu NMR spectrum is observed (see the top panel in Fig.~\ref{fig:EuNMR} where the simulated spectum shown by the red line is the same as the case of $H$ = 0). 
    Since the effective field at the Eu site is given by the vector sum of  $\bf{B}_{\rm int}^{\rm Eu}$ and $\bf{H}$, i.e., $|$$\bf{B}_{\rm eff}$$|$ = $|$$\bf{B}_{\rm int}^{\rm Eu}$ + $\bf{H}$$|$,   the resonance frequency is expressed for $H$ $\perp$ $<$$\mu$$>$  as $f$ =  $\frac{\gamma_{\rm N}}{2\pi}$$\sqrt {(B_{\rm int}^{\rm Eu})^2+H^2}$. 
    For our applied field range where $B_{\rm int}^{\rm Eu}$ $>>$ $H$, any shift in the resonance frequency due to $H$ would be small, as observed.
    Thus, we conclude that $H$ is perpendicular to $B_{\rm int}^{\rm Eu}$ and thus to the ordered Eu moments. 
 
    In the case of ${\bf H}$ applied parallel to the $ab$ plane, on the other hand, each line broadens and exhibits a typical two-horn structure expected for an incommensurate planar helical structure as shown in the middle panel of Fig.~\ref{fig:EuNMR}.  
   In fact, the observed spectrum at $H$ = 1~T is well reproduced by  a calculated spectrum for an incommensurate helical AFM state shown by the red line.
   The inset of the middle panel of  Fig.\ \ref{fig:EuNMR} shows the external field dependence of the amount of the splitting of the central transition line ($\Delta f$) of the $^{153}$Eu ZFNMR spectra. 
   The $\Delta f$ increases with increasing $H$.
   Since the peak positions of the two-horn shape of the spectrum are given by $B_{\rm eff}$ = $B_{\rm int}$ $\pm$ $H$, the $\Delta f$ is proportional to $H$ according to $\Delta f$ = 2$H$$\gamma_{\rm N}$/(2$\pi$). 
   As shown by the solid line in the inset, the $H$ dependence of $\Delta f$ is well reproduced by this relation.
    Thus these NMR results are consistent with  an incommensurate helical spin structure with the ordered moments aligned along the $ab$ plane as  reported from the ND  \cite{Reehuis1992} and magnetization \cite{Sangeetha2016} measurements. 
   The observed $ab$-plane alignment of the ordered moments is also consistent with the prediction of the moment alignment from magnetic dipole interactions between the Eu spins \cite{Johnston2016}.

 \subsection{B. $^{31}$P NMR spectrum}   

\begin{figure}[tb]
\includegraphics[width=7.0cm]{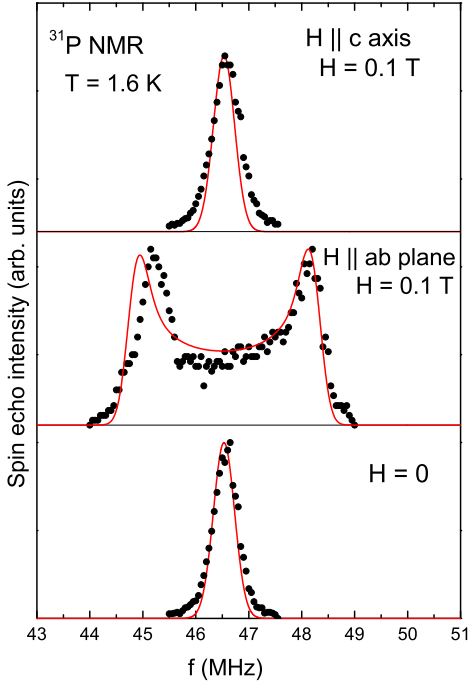} 
\caption{(Color online)  $^{31}$P-NMR spectra at $T$ = 1.6 K in the AFM state for EuCo$_2$P$_2$ in zero magnetic field (bottom panel), and under magnetic fields parallel to the $ab$ plane (middle panel) and parallel to the $c$ axis (top panel).   
  The red lines are the calculated $^{31}$P-NMR spectra. 
 }
\label{fig:PZFNMR}
\end{figure}

\begin{figure}[tb]
\includegraphics[width=7.0cm]{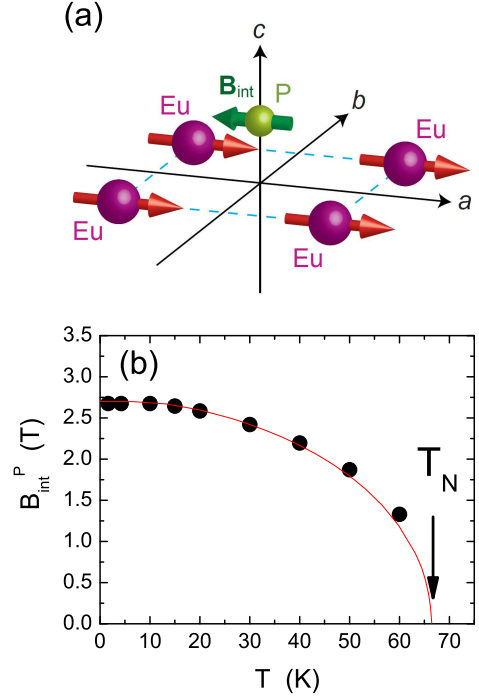} 
\caption{(Color online)  (a)   Coordination of nearest-neighbor Eu sites around a P site. The arrows on the Eu and P atoms indicate the directions of the Eu ordered  moments and the internal magnetic induction at the P site, respectively.
(b) Temperature dependence of $B_{\rm int}^{\rm P}$. The solid curve is the Brillouin function with $J$ = $S$ = 7/2. 
}
\label{fig:PNMR}
\end{figure}

  The incommensurate planar helix structure is also clearly evidenced by $^{31}$P NMR measurements.
    The bottom panel of Fig.~\ \ref{fig:PZFNMR} shows the $^{31}$P ZFNMR spectrum at 1.6~K in the AFM state, where the red line is the fit with the parameter $|B_{\rm int}^{\rm P}|$  = 2.69~T.
 For $I$ =1/2, a single NMR line is expected and observed, because of no quardupole interaction.
       When  ${\bf H}$ is applied along the $c$ axis, almost no change of the spectrum is observed as typically shown in the top panel of Fig.~\ref{fig:PZFNMR} where $H$ = 0.1~T.
    For comparison, we show the red line calculated for $H$ = 0. 
   This indicates that $\bf H$ is perpendicular to $\bf  B_{\rm int}$ at the P site. 
    On the other hand, when $\bf H$ = 0.1~T is applied parallel to the $ab$ plane, similar to the case of the $^{153}$Eu NMR spectrum,  the line exhibits a characteristic two-horn shape, again expected for the incommensurate planar helix AFM state.

    As discussed for EuGa$_4$ (Ref. \cite{Yogi2013}) and EuCo$_2$As$_2$ (Ref. \cite{Ding2017}), the direction of $B_{\rm int}^{\rm P}$  is antiparallel to the Eu ordered moments in the case where the Eu ordered moments are ferromagnetically aligned in the Eu plane, as shown in Fig.~\ref{fig:PNMR}(a).
    Therefore, one can expect almost no change of the $^{31}$P NMR spectrum when $H$ is perpendicular to the Eu ordered moment, as observed in the $^{31}$P NMR spectrum for $H$ $\parallel$ $c$ axis.
    On the other hand,  if one applies $H$ $\parallel$ $ab$ plane, a splitting of the $^{31}$P ZFNMR spectrum is expected similar to the case of the  $^{153}$Eu ZFNMR spectrum. 
    The red line in the middle panel of Fig.~\ref{fig:PZFNMR} is the calculated spectrum of $^{31}$P NMR for the planar helix AFM structure under $H$ = 0.1~T, which reproduces the observed spectrum very well.

\begin{figure}[tb]
\includegraphics[width=7.0cm]{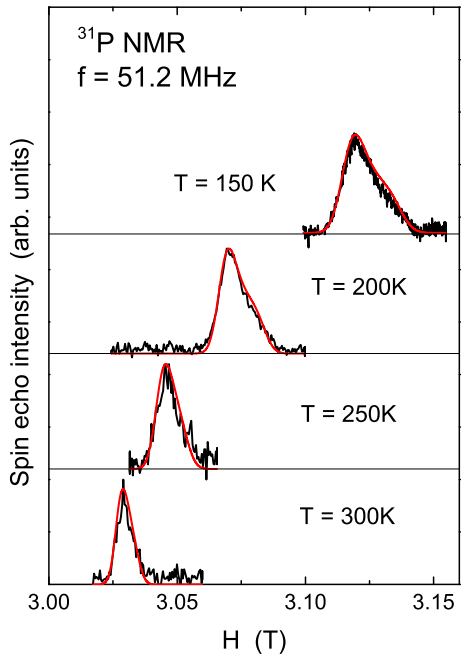} 
\caption{(Color online)  Field-swept $^{31}$P-NMR spectra of EuCo$_2$P$_2$ (powder sample) measured at $f$ = 51.2 MHz.  
      The red lines are simulated spectra with hyperfine anisotropy.  }
\label{fig:PNMR2-1}
\end{figure}

\begin{figure}[b]
\includegraphics[width=7.0cm]{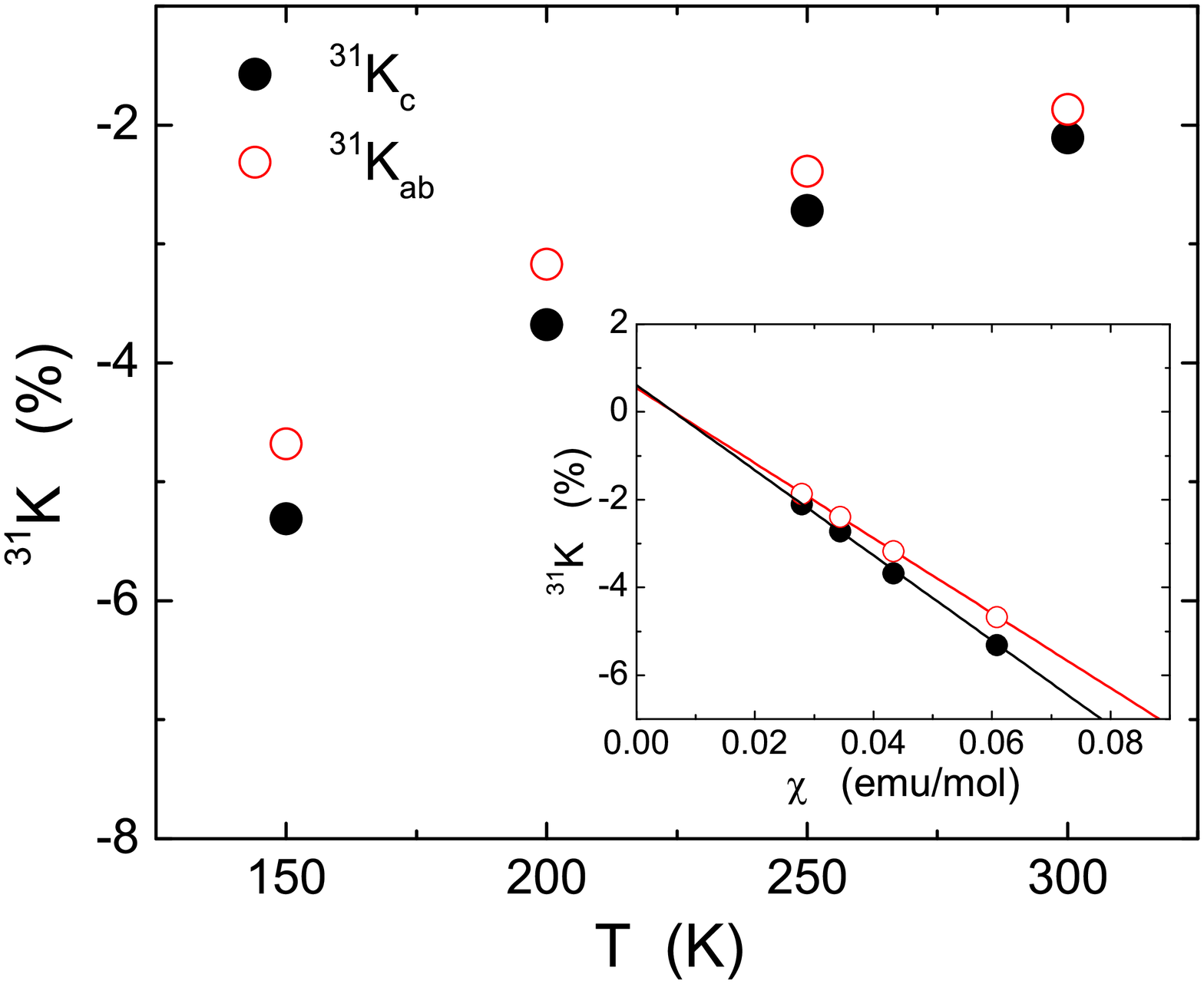} 
\caption{(Color online)   Temperature dependences of the $^{31}$P Knight shifts $^{31}K$ in the paramagnetic state. 
 The inset shows  $K(T)$ versus magnetic susceptibility $\chi(T)$.  The solid  lines are linear fits. 
 }
\label{fig:PNMR2-2}
\end{figure}

    The $T$ dependence of the $^{31}$P ZFNMR spectrum was measured up to 60 K.  
   With increasing $T$, the spectra shift to lower frequency due to reduction of the internal magnetic induction  $|B_{\rm int}^{\rm P}|$ which decreases from 2.69~T at 1.6~K to 1.33~T at 60~K. 
     The $T$ dependence of $|B_{\rm int}^{\rm P}|$  is shown in Fig.~\ref{fig:PNMR}(b), which is the $T$ dependence of  the order parameter of the planar helix AFM state, and is well reproduced by a Brillouin function which was calculated based on the Weiss molecular field model with $J$ = $S$ = 7/2, $T_{\rm N}$ = 66.5 K and $B_{\rm int}^{\rm P}$ = 2.69~T at $T$=1.6~K [solid curve in Fig.~\ref{fig:PNMR}(b)]. 
   This indicates that  the magnetic state of the Eu ions is well explained by the local moment picture although the system is metallic as determined from electrical resistivity measurements \cite{Sangeetha2016}.

  Now we discuss our  NMR data for the PM  state.
    Figure \ref{fig:PNMR2-1} shows the  temperature dependence of the field-swept $^{31}$P NMR spectra for powdered single crystals and $T$ = 150 K to 300 K.
     With decreasing $T$, the peak position shifts to higher magnetic field and the line becomes broader and asymmetric due to anisotropy in the  Knight shifts in the powder sample.
     We determine the NMR shifts for $H$ parallel to the $c$ axis ($^{31}K_c$) and parallel to the $ab$ plane ($^{31}K_{ab}$) from fits of the spectra as shown by the red lines. 
    The $T$ dependences of $^{31}K_c$ and $^{31}K_{ab}$ are shown in Fig. \ref{fig:PNMR2-2}.
     The hyperfine coupling constants  $A^{\rm P}$ of $^{31}$P surrounded by Eu$^{2+}$ ions can be estimated 
from the slopes of $K$-$\chi$ plots with the relation   
\begin{equation}
   A   = \frac{N_{\rm A}}{Z}\frac{K(T)}{\chi(T)},
\label{eqn:Kchi}
\end{equation}
where  the $\chi(T)$  data are from Ref. \cite{Sangeetha2016}, and ${N_{\rm A}}$ is Avogadro's number and $Z=4$ is the number of nearest-neighbor (NN) Eu$^{2+}$  ions around a P atom. 
   Here we assume the hyperfine field at the P sites mainly originates from the NN Eu spins.     
      As shown in the inset of Fig.\ \ref{fig:PNMR2-2}, both $^{31}K_c$ and $^{31}K_{ab}$ vary linearly with $\chi$. 
      From the respective slopes, the hyperfine coupling constants $A^{\rm P}_c $ and  $A^{\rm P}_{ab}$  are estimated to be $-$1.43 $\pm$ 0.10 and $-$1.23 $\pm$ 0.09 kOe/$\mu_{\rm B}$/Eu, respectively.
 
    With the value of $|B_{\rm int}^{\rm P}|$  = 2.69~T,  the Eu ordered moment is estimated  to be $<$$\mu$$>$ = 5.46 $\mu_{\rm B}$ at 1.6 K using the relation $|B_{\rm int}^{\rm P}|$ = 4$A^{\rm P}_{ab}$$<$$\mu$$>$. 
    This value is smaller than 6.9~$\mu_{\rm B}$/Eu reported from the ND study \cite{Reehuis1992}. 
    The difference may suggest that the estimated  $A^{\rm P}_{ab}$ in the PM state is slightly greater than that in the AFM state. 
    This would be possible if one takes  finite contributions to hyperfine fields from the next-nearest-neighbor (NNN) Eu spins on the next layer into consideration.
     Since in the AFM state  the direction of NNN Eu spins is antiparallel to that of the NN Eu spins, one expects a positive hyperfine field at the P sites which cancels part of the negative hyperfine field produced by the NN Eu spins.  
     Assuming  $<$$\mu$$>$ = 6.9~$\mu_{\rm B}$ (Ref. \cite{Reehuis1992}), we thus estimate a $\sim$ 21 \% net additional contribution to the hyperfine field from the NNN Eu spins.

 \subsection{C. $^{59}$Co NMR spectrum}   

\begin{figure}[tb]
\includegraphics[width=7.0cm]{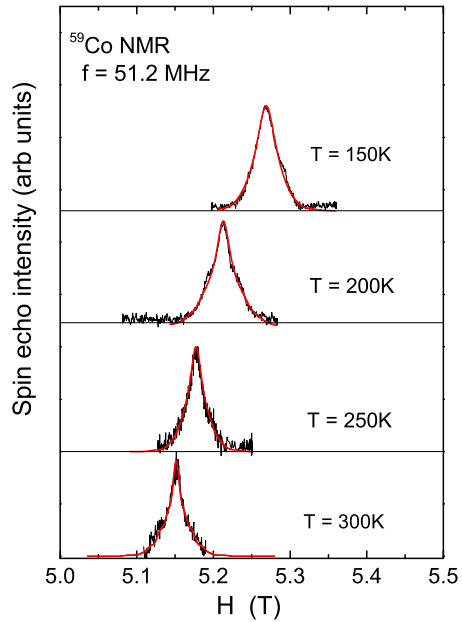} 
\caption{(Color online) Field-swept $^{59}$Co-NMR spectra in the paramagnetic state of EuCo$_2$P$_2$ (powder sample) measured at $f$ = 51.2 MHz.  
 }
\label{fig:CoNMR1}
\end{figure}

\begin{figure}[tb]
\includegraphics[width=7.0cm]{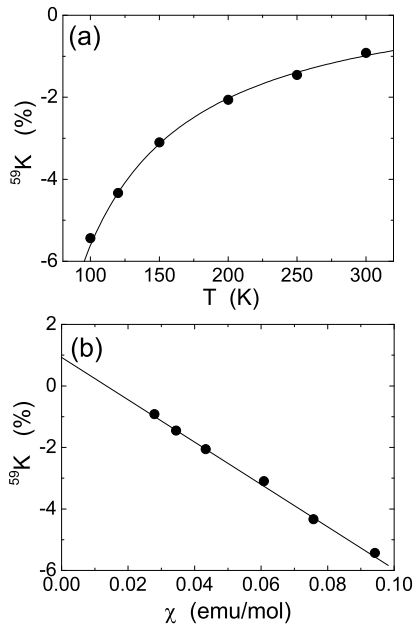} 
\caption{(Color online) (a) Temperature dependence of $^{59}$Co Knight shift $^{59}K$ in the paramagnetic state.
     The solid line is fit to the data by a Curie-Weiss law. 
 (b)   $K(T)$ versus magnetic susceptibility $\chi(T)$.  The solid  line is linear fit. 
}
\label{fig:CoNMR2}
\end{figure}

    Figure \ref{fig:CoNMR1} shows the  temperature dependence of the field-swept $^{59}$Co NMR spectra in the PM state of a powder sample where the spectra are seen to broaden with decreasing $T$. 
    Although one expects a central transition line with three satellite lines on both sides for $I$ = 7/2 nuclei,    
    the observed spectra do not show the seven peaks but rather exhibit a single broad line due to inhomogeneous magnetic broadening. 
    Since the powder sample consists of grains with randomly oriented crystal axes, the spectra are powder patterns. 
    From the fitting of the spectra shown by red lines which are calculated from the nuclear spin Hamiltonian with the Zeeman interaction much greater than the quadrupole interaction, we estimate  $\nu_{\rm Q} \sim 0.25 $ MHz which is nearly independent of $T$.  
    The broadening of the spectra with lowering $T$ originates from magnetic broadening. 
     The $T$ dependence of the NMR shift $^{59}K$ determined from the peak position of the spectrum is shown in Fig. \ref{fig:CoNMR2}(a), where we fit the data with the Curie-Weiss law $\frac{C}{T-\theta_{\rm p}}$. 
     The solid line is a fit with $C$ = -533(13) $\% \rm K$ and $\theta_{\rm p}$ = 18(3) K for $^{59}K$.  
     The value $\theta_{\rm p}$ = 18(3)  K  is the same within the error as the powder averaged value obtained for a single crystal of EuCo$_2$P$_2$ from $\chi(T)$ measurements \cite{Sangeetha2016}.
   The positive value of $\theta_{\rm p}$ indicates predominant ferromagnetic (FM) exchange interactions between the Eu spins. 
   This is consistent with the in-plane FM exchange interactions responsible for the planar helix AFM structure.
    The hyperfine coupling constant  $A^{\rm Co}$ for $^{59}$Co surrounded by $Z=4$ Eu$^{2+}$ ions is also estimated 
from the slope of $K$-$\chi$ plot in Fig. \ref{fig:CoNMR2}(b) with Eq.~(\ref{eqn:Kchi}).    
   We thus estimate $A^{\rm Co}$ = ($-$0.98 $\pm$ 0.09) kOe/$\mu_{\rm B}$/Eu.
   This value is much smaller than a typical value $A$  = $-$105  kOe/$\mu_{\rm B}$ for Co $3d$ electron core polarization \cite{Freeman1965}.  
   This indicates that the hyperfine field at the Co site originates from the transferred hyperfine field produced by the Eu spins and that no $3d$ spins on the Co sites contribute to the magnetism of EuCo$_2$P$_2$.

\begin{figure}[tb]
\includegraphics[width=7.0cm]{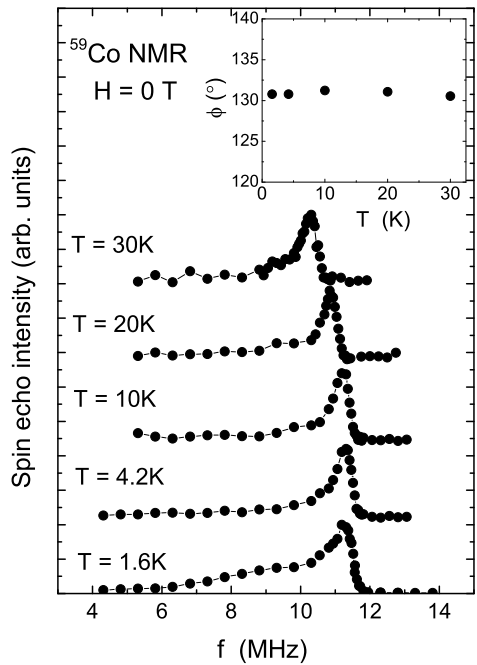} 
\caption{(Color online)  $^{59}$Co zero-field NMR spectra in the AFM state in zero magnetic field at the indicated temperatures. The inset shows the temperature dependence of the turn angle $\phi$.  }
\label{fig:CoZFNMR}
\end{figure}

  We now consider the influence of the planar helix AFM state on the Co NMR data.
  We tried to detect the signals in the AFM state using a single crystal but we could not find any. 
    Then we used powder samples for which we succeeded in observing the $^{59}$Co ZFNMR spectrum up to 30 K as shown in Fig.~\ref{fig:CoZFNMR}. 
    From the peak position of the spectrum, the internal magnetic induction at the Co site at 1.6~K is estimated to be  $|B_{\rm int}^{\rm Co}|$  = 11.3~kOe which decreases to 10.2~kOe at 30~K. 
    According to  the analysis performed in EuCo$_2$As$_2$ \cite{Ding2017}, one can estimate the AFM propagation vector in the incommensurate state of EuCo$_2$P$_2$ based on  the estimated values of $|B_{\rm int}^{\rm Co}|$ and  $A^{\rm Co}$. 
    Here we present a similar discussion as in the previous paper \cite{Ding2017}.

\begin{figure}[tb]
\includegraphics[width=8.5cm]{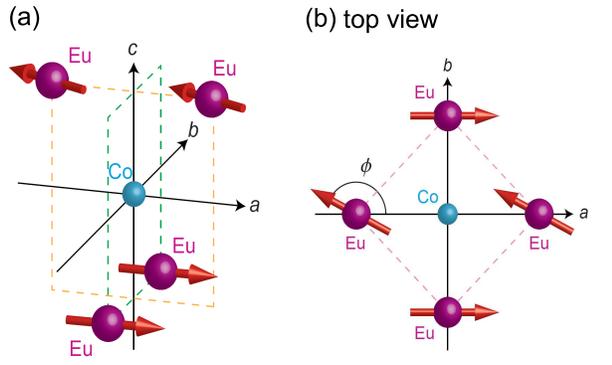} 
\caption{(Color online)  (a) Coordinations of nearest-neighbor Eu sites around a Co site. The arrows on the Eu atoms indicate the ordered magnetic moments. 
(b) Top view of the coordination of nearest-neighbor Eu sites around Co site. The magnetic moment turn angle between adjacent magnetic layers is $\phi$.   }
\label{fig:CoZFNMR2}
\end{figure}

    In an incommensurate helical AFM state, $|B_{\rm int}|$  at the Co site appears only in the $ab$ plane when the Eu ordered moments lie in the $ab$ plane and is expressed by  \cite{Ding2017}
\begin{eqnarray}
 B_{\rm int}^{\rm Co} =  2 \langle \mu \rangle A^{\rm Co}\sqrt{2+2\cos\phi},
\label{eqn:Co_internal}
\end{eqnarray}
where $\phi$ is the turn angle  along the $c$ axis between the Eu ordered moments in adjacent Eu planes, which characterizes the helical structure. 
    In the case of  $\phi$ = $\pi$ corresponding to a collinear AFM state,  $ B_{\rm int}^{\rm Co} $ is zero due to a cancellation of the internal magnetic induction from the four nearest-neighbor Eu ordered moments.  
   On the other hand, if $\phi$ deviates from $\pi$ corresponding to a helical state, one can expect a finite $ B_{\rm int}^{\rm Co}$ [see Fig.\ \ref{fig:CoZFNMR2}(a)].  
   Thus the observation of the finite $ B_{\rm int}^{\rm Co}$  is direct evidence of the planar incommensurate helix AFM state in EuCo$_2$P$_2$.
   Furthermore, using Eq.~(\ref{eqn:Co_internal}), we can determine the AFM propagation vector {\bf k} = (0, 0, $k$)2$\pi$/$c$, where $c$ is the $c$-axis lattice parameter of the body-centered tetragonal Eu sublattice.  
    Since the distance $d$ along the $c$ axis between adjacent layers of FM-aligned Eu moments is $d$ = $c$/2,  the turn angle between the ordered moments in adjacent Eu layers is $\phi$ = $kd$, as shown in Fig.\ \ref{fig:CoZFNMR2}(b).
    Using $\langle$$\mu$$\rangle$ = 6.9(1) $\mu_{\rm B}$ \cite{Reehuis1992},  $A^{\rm Co}$ = $-$0.98(9)  kOe/$\mu_{\rm B}$/Eu and  $B_{\rm int}^{\rm Co}$ = 11.3(1) kOe, the turn angle $\phi$ is estimated to be 131$\pm$16$^{\circ}$ corresponding to a helix wave vector  {\bf k} = (0, 0, 0.73 $\pm$ 0.09)2$\pi$/$c$.
   This value of {\bf k}  is slightly smaller than than  ${\bf k}$ = (0, 0, 0.852)2$\pi$/$c$ obtained from the ND data \cite{Reehuis1992} and (0, 0, 0.88)2$\pi$/$c$ estimated from the $\chi$ data \cite{Sangeetha2016} on  EuCo$_2$P$_2$, and is close to  {\bf k} = (0, 0, 0.73)2$\pi$/$c$ determined by the NMR data in EuCo$_2$As$_2$ \cite{Ding2017}. 
  The origin of the small difference in {\bf k}  between the NMR and ND (and $\chi$) data is not clear, but it could be explained, e.g., if one would take other small contributions to the hyperfine field at the Co site from the NNN Eu spins.

   The asymmetric  shape of the observed $^{59}$Co ZFNMR spectrum originates from a distribution of  the internal field at the Co sites. 
   The $^{153}$Eu ZFNMR lines are sharp as seen in Fig.~\ref{fig:EuNMR},  indicating homogeneous Eu ordered moments. 
   The very sharp $^{31}$P ZFNMR line also indicates that the direction of the Eu moments in each ferromagnetic Eu plane is relatively uniform. 
   Therefore, the low-frequency tail of the Co ZFNMR spectrum suggests a distribution of the turn angle $\phi$, i.e., the AFM propagation vector ${\bf k}$. 
     Using the values of the internal field distribution of the Co site from 0.6 T (6 MHz) to 1.2 T (12 MHz), the distribution of the turn angle $\phi$ is estimated to be  from 156 to 131$^{\circ}$. 
    This corresponds to a change in ${\bf k}$ from (0, 0, 0.86)2$\pi$/$c$ to (0, 0, 0.73)2$\pi$/$c$.  
     It is worth mentioning that the NMR technique determines not only the AFM propagation vector but also its distribution.

    Finally we discuss the temperature dependence of the turn angle $\phi$.    
 Assuming the temperature dependence of the Eu ordered moments $\langle \mu \rangle$ is described by the temperature dependence of $B_{\rm int}^{\rm P}$, one can estimate $\phi$ at each temperature based on the  temperature dependence of  $B_{\rm int}^{\rm Co}$ using Eq. (\ref{eqn:Co_internal}). 
    As shown in the inset of Fig.\ \ref{fig:CoZFNMR}, $\phi$ is nearly independent of temperature up to 30 K. 
    According to the ND measurements, on the other hand, $\phi$ changes from 150$^{\circ}$ at 64 K just below $T_{\rm N}$ to 153$^{\circ}$ at 15 K \cite{Reehuis1992}. 
   Thus the small 2\% change in $\phi$ observed in the ND measurements may occur at temperatures higher than 30 K.

   \section{IV. Summary and concluding remarks}

     We have carried out  $^{153}$Eu, $^{31}$P and $^{59}$Co NMR measurements on the helical antiferromagnet EuCo$_2$P$_2$ with $T_{\rm N}$ = 66.5 K. 
    The external magnetic field dependence of  $^{153}$Eu and $^{31}$P NMR spectra for  single crystalline EuCo$_2$P$_2$ clearly evidenced the incommensurate helical AFM structure. 
   The AFM propagation vector characterizing the incommensurate helical AFM state was determined  to be {\bf k} = (0, 0, 0.73 $\pm$ 0.09)2$\pi$/$c$ from the internal magnetic field at the Co site obtained by $^{59}$Co NMR under zero magnetic field. 
    The AFM propagation vector is revealed to be nearly independent of temperature up to 30 K, indicating that the small change in the propagation vector observed by ND measurements may occur at temperatures higher than 30 K. 
 
      As described in \cite{Ding2017}, our NMR approach can be used to study in detail the magnetism originating from the Eu spins in the Co-substituted iron-pnictide high-$T_{\rm c}$ superconductor Eu(Fe$_{1-x}$Co$_x$)$_2$As$_2$.
       The $x$ = 0 compound,  EuFe$_2$As$_2$, exhibits the stripe-type AFM order at 186~K due to the Fe spins. 
   At the same time, the Eu$^{2+}$ moments order antiferromagnetically below 19 K  with  the A-type AFM structure where the Eu ordered moments are FM aligned in the $ab$ plane but the moments in adjacent layers along the $c$ axis are antiferrmagnetically aligned \cite{Jeevan2008}.
     With substitution of Co for Fe  in Eu(Fe$_{1-x}$Co$_x$)$_2$As$_2$, the magnetic structure of the Eu$^{2+}$ spins changes from the A-type AFM order in the $x$ = 0 compound, to the A-type canted AFM structure  at intermediate Co doping levels around $x \sim$ 0.1, and then  to the FM order along the $c$ axis at $x$ $\sim$ 0.18 where superconductivity (SC) appears  below $T_{\rm c}$ $\sim$10~K in the range $x$ = 0.1 to 0.18  \cite{Jin2016}.  
   Thus it is important to understand the magnetism originating from the Eu and Fe spins and also the SC properties from a microscopic point of view. 
    Our approach based on the NMR technique provides an important avenue  to study the origin of the coexistence of SC and magnetism in Eu(Fe$_{1-x}$Co$_x$)$_2$As$_2$ SCs.

  \section{V. Acknowledgments} 
   We thank H. Uehara and F. Kubota for assistance with the experiments
    The research was supported by the U.S. Department of Energy, Office of Basic Energy Sciences, Division of Materials Sciences and Engineering. Ames Laboratory is operated for the U.S. Department of Energy by Iowa State University under Contract No.~DE-AC02-07CH11358.
    Part of the work was supported by  the Japan Society for the Promotion of Science KAKENHI : J-Physics (Grant Nos. JP15K21732, JP15H05885, and JP16H01078).
    N. H. also thanks the KAKENHI : J-Physics for financial support to be a visiting scholar at the Ames Laboratory.

\end{document}